\documentclass[11pt,a4paper]{article}
\usepackage{jheppub} 

\usepackage[notref,notcite]{}

\def \mvec #1{\mbox{\boldmath{${#1}$}}}
\def \msvec #1{\mbox{\scriptsize \boldmath{${#1}$}}}

\newcommand{\rmd}{{\rm d}}

\title{Walking in the 3-dimensional large $N$ scalar model}

\author[a,b]{Sinya\ Aoki,}
\author[c]{Janos \ Balog,}
\author[d]{Peter \ Weisz}

\affiliation[a]{Yukawa Institute for Theoretical Physics, Kyoto University, \\
Kitashirakawa Oiwakechou, Sakyo-ku, Kyoto 606-8502, Japan}
\affiliation[b]{Center for Computational Sciences, University of Tsukuba 305-8577, Japan}
\affiliation[c]{Institute for Particle and Nuclear Physics, 
Wigner Research Centre for Physics, \\
MTA Lend\"ulet Holographic QFT Group,
1525 Budapest 114, P.O.B.\ 49, Hungary}
\affiliation[d]{Max-Planck-Institut f\"ur Physik, 80805 Munich, Germany}

\emailAdd{saoki@yukawa.kyoto-u.ac.jp}
\emailAdd{balog.janos@wigner.mta.hu}
\emailAdd{pew@mpp.mpg.de}

\abstract{The solvability of the three-dimensional O($N$) scalar field theory in the large $N$
limit makes it an ideal toy model exhibiting "walking" behavior, expected in some
SU($N$) gauge theories with a large number of fermion flavors. We study the model
using lattice regularization and show that when the ratio of the particle mass
to an effective 4-point coupling (with  dimension mass) is small, the beta function
associated to the running 4-point coupling is "walking". We also study lattice
artifacts and finite size effects, and find that while the former can be sizable
at realistic correlation length, the latter are under control already at lattice
sizes a few ($\sim$3) correlation lengths. We show the robustness of the walking
phenomenon by showing that it can also be observed by studying physical
observables such as the scattering phase shifts and the mass gap in finite volume.}


\subheader{\hfill YITP-14-53, MPP-2014-296}

\begin{document} 

\maketitle

\section{Introduction}
\label{intro}
The Standard Model (SM) is a truly remarkable effective theory,
presently in agreement with all available electro-weak data
(apart from a few tensions). 
However, in search of a more fundamental theory,
it was proposed\cite{Weinberg, Susskind}
that a higher gauge theory, called technicolor (TC)
(for a detailed review see \cite{HiSi}),
could dynamically explain the electro-weak scale and the
origin of masses of the gauge bosons.

The original TC models, however, had the difficulty that they could not give quark masses, and
models proposed to overcome this, 
so-called extended TC (ETC) models \cite{DiSu,EiLa}, 
have a tension between a suppression of effects caused by flavor-changing neutral currents
and a large top quark mass.  In addition, precise electroweak measurements give strong constraints on the dynamics of TC models\cite{PeTa}.
In order to overcome these difficulties,
a class of asymptotically free 
so-called walking TC models, for which the 
characteristic dynamics changed very slowly 
in the low energy regime, was proposed\cite{Holdom, YaBaMa, ApWi}.
This could occur if typical running couplings
came close to conformal fixed points where the associated
beta-functions became zero \cite{BaZa}.

The walking phenomena in asymptotically free theories is a 
non-perturbative phenomenon and requires corresponding 
investigations such as numerical simulations of the 
lattice regularized theory.  
This program has  been actively and systematically pursued recently 
\cite{AFN}  given the improvement of lattice technology to simulate 
dynamical fermions and with the motivation of finding an 
alternative to SUSY to describe physics beyond the SM.
The most popular models in 4-dimensions studied so far
are SU($N$) gauge theories with a large number $N_f$ of fermion 
flavors, or with fermions in higher representations.
See recent reviews\cite{Fleming, Pallante, DelDebbio, Neil, Nogradi,  Giedt, Kuti, Itou} for  references to these activities.
For gauge group SU($N$) it is probable that walking sets 
in when $N_f$ exceeds a certain value (still less than
$11N/2$ to maintain asymptotic freedom).
Although there are reports that SU(3) gauge theory with
$N_f=8$ show walking behavior \cite{Nagai}, 
there is still some debate \cite{HCPS}. 
Unambiguous identification of walking phenomena is 
in these models extremely difficult and CPU intensive.
This is not only due to the problem of 
efficiently simulating dynamical fermions, 
but also to the need to control systematic errors arising 
from finite ultra-violet cutoff (finite lattice spacings)
and from finite volume effects \cite{GiWe,CDGK}. 

It is thus advantageous to have simple toy models 
at hand which show walking and where the systematic errors 
mentioned above can be investigated in detail. 
There are as yet not many toy models where walking phenomena
have been studied. One recent example suggested 
by Nogradi \cite{Nogradi2} and studied numerically 
by de Forcrand, Pepe and Wiese \cite{deFPW} 
is the non-linear O(3) model in two dimensions 
with a $\theta$ term. 
In this model it was established that
walking appears when the continuous parameter
$\theta$ is close to the value $\pi$. 

In this paper we investigate properties of another simple
model, the $N$-component scalar field theory in three
dimensions in the large $N$ limit.
This model exhibits walking phenomena when a continuous
parameter $\alpha$, the ratio of renormalized mass to an effective
4-point coupling (with dimension mass), is very small.
Other aspects of this model have been frequently studied
in the literature (see the review \cite{MoZi}).
In this toy model many features can be 
studied analytically, for example the systematic sources 
of error in lattice simulations mentioned above.

The material presented in this paper is as follows.
In sect.2 we define the model with a particular lattice
regularization and discuss its large $N$ approximation.
In the simplest scenario the lattice model has
two phases, one where the O($N$) symmetry is retained
and one where this symmetry is broken spontaneously. 
Section 3 discusses renormalization and various continuum
limits which can be obtained by approaching the critical
surface in various ways.

In section 4 we introduce a dimensionless running coupling 
defined naturally through the connected 4-point function
in the symmetric phase and discuss its behavior in the 
various continuum limits. One case, which we call IIA, 
exhibits walking for small values of $\alpha$, 
and in a typical case we investigate the lattice artifacts.
Section 5 discusses the scattering phase shift in the
various continuum limits, as an on-shell dynamical quantity 
which also shows walking for case IIA.

We end our presentation with a summary of our results. 
In addition there are two appendices. 
Appendix A discusses the 
parameters of the resonance present in the broken phase
of the lattice regulated theory. In Appendix B we
introduce a coupling running with the volume defined
in terms of the mass gap in a finite volume with periodic
boundary conditions. 
This coupling also exhibits walking in case IIA.

\section{Large $N$ expansion of the lattice model}
\label{largen}
In this paper, we consider an $N$ component scalar model in 3 dimensions, defined by the action
\begin{eqnarray}
S &=& \int \mathrm{d}^3 x \left[ \frac{1}{2}\partial^\mu \phi (x)\cdot \partial_\mu \phi (x) + N V_{\rm cont}\left(\frac{\phi^2(x)}{N} \right)
\right] ,
\label{eq:action_cont}
\end{eqnarray}
where $\phi^i(x)$ is an $N$ component scalar field with $i=1,2,\cdots,N$,   $(\ \cdot \ )$ indicates an inner product of $N$ component vectors such that $\phi^2(x) \equiv \phi(x)\cdot\phi(x) =\sum_{i=1}^N \phi^i(x)\phi^i(x)$, and $V_{\rm cont}$ is a potential term, which takes the form 
$
V_{\rm cont}(s) = \frac{m_0^2}{2} s + \frac{g_4}{4!} s^2 + \frac{g_6}{6!} s^3,
$
where $m_0$ is the bare scalar mass, $g_{k}$ is the bare coupling constant of the $\phi^{k}$ interaction, whose canonical dimension  is $3-k/2$.

We put  the above model on a 3 dimensional Euclidean lattice with periodic boundary conditions in all directions.
Let us define the partition function with a source as
\begin{eqnarray}
Z(J) &=& \int \left[{\cal D}\Phi\right]\exp\left\{\sum_{\msvec n}\left[
\sum_{\mu=0}^2 \Phi_{\msvec n}\cdot \Phi_{{\msvec n}+\hat{\mu}}
-N U\left(\frac{1}{N}\Phi^2_{\msvec n}\right)
+J_{\msvec n}\cdot \Phi_{\msvec n}\right]\right\},
\end{eqnarray}
where  $\mvec n =(n_0,n_1,n_2)$ with $n_i \in \mvec Z$, $\hat \mu$ is the unit vector in the $\mu$-th direction, and $
\left[{\cal D}\Phi\right]=\displaystyle \prod_{{\msvec n},i}{\rm d}\Phi^i_{\msvec n}
$
is the measure.
The periodic boundary condition is implemented by 
\begin{eqnarray}
 \Phi^i_{\msvec n + L_0\hat{0}}= \Phi^i_{\msvec n + L_1\hat{1}} &=&  \Phi^i_{\msvec n + L_2\hat{2}}=\Phi^i_{\msvec n},
\end{eqnarray}
where ${\mvec L}=(L_0,L_1,L_2)$  are large but finite integers.  We can obtain correlation functions of $\Phi$'s  from $Z(J)$,
by taking its derivative  with respect to $J$'s and then setting $J_{\msvec n}^i \rightarrow J_0^i\equiv \sqrt{N} H \delta^{iN}$, where the constant external field  $H$ coupled only to $N$-th component of $\Phi$ is introduced to handle a possible spontaneous breaking of O($N$) symmetry.

A potential term on the lattice takes the form
\begin{equation}
U(S)=\frac{R}{2}S+\frac{U}{4!}S^2+\frac{G}{6!}S^3+\sum_{k\ge 4} G_k S^k,  
\end{equation}
where dimensionless field $\Phi_{\msvec n}$ and   
parameters $R, U, G$  are related to the continuum ones via the lattice spacing $a$ as
\begin{equation}
\Phi_{\msvec n}= \sqrt{a} \phi( x=\mvec n a), \quad R = m_0^2 a^2+6, \quad U = g_4 a, \quad G = g_6,
\end{equation}
while $G_{k\ge 4}$ on the lattice does not have an counterpart in the formal continuum theory (\ref{eq:action_cont}).
Of course, these relations are classical and thus are modified after renormalization.
An existence of the lower bound of the potential $U(S)$ requires that non-zero coefficient of the largest power of $S$ must be positive. Near the continuum limit this 
requires that $ G\ge 0$, and if  $G=0$, then $U\ge 0$.
In this paper, for distinction between dimensionless and dimensional quantities, we use capital letters for dimensionless quantities as much as possible.

\subsection{Large $N$ saddle point expansion}
Preparing the large $N$ expansion, we insert into the path integral
\begin{equation}
1\equiv\int\left[{\cal D}\Omega\right] \prod_{\msvec n}\delta\left(\Omega_{\msvec n}
-\frac{1}{N}\Phi^2_{\msvec n}\right)={\cal N}_1
\int\left[{\cal D}\Lambda\right]\left[{\cal D}\Omega\right]
\exp\left\{-N\sum_{\msvec n}i\Lambda_{\msvec n}\left(\frac{1}{N}
\Phi^2_{\msvec n}-\Omega_{\msvec n}\right)\right\},
\end{equation}
where ${\cal N}_1=\displaystyle \left(\frac{N}{2\pi} \right)^{V}$ with $V=L_0L_1L_2$, so that $\Phi$-integral becomes Gaussian and can be performed. We then obtain
\begin{equation}
Z(J)={\cal N}_2 \int\left[{\cal D}\Lambda\right]\left[{\cal D}\Omega\right]
\exp[-N S_{\rm eff}(\Lambda,\Omega,J)],\quad
{\cal N}_2 ={\cal N}_1 (2\pi)^{N V /2}
\label{master2}
\end{equation}
where the effective action is given by
\begin{equation}
S_{\rm eff}(\Lambda,\Omega, J)=\sum_{\msvec n}\left[ U(\Omega_{\msvec n}) -i\Lambda_{\msvec n}\Omega_{\msvec n}\right]-\frac{1}{2N}\sum_{{\msvec n}{\msvec m},i}
J^i_{\msvec n} \left(D^{-1}[\Lambda]\right)_{\msvec {nm}} J^i_{\msvec m} +\frac{1}{2}{\rm Tr} \ln D[\Lambda] ,
\label{Seff}
\end{equation}
and the matrix $D[\Lambda_{\msvec n}] $ here acts on an arbitrary vector $F$ as
\begin{eqnarray}
\left( D[\Lambda]  F \right)_{\msvec n} &=& 2(i\Lambda_{\msvec n}-3) F_{\msvec n} -(\nabla^2 F)_{\msvec n}, \quad
(\nabla^2 F)_{\msvec n}\equiv \sum_\mu\left(F_{\msvec n +\hat\mu} + F_{\msvec n - \hat\mu} -2 F_{\msvec n}\right).
\label{Mmatrix}
\end{eqnarray}

The large $N$ limit corresponds to the saddle point, determined  by the saddle point equations,
\begin{eqnarray}
 \frac{\partial S_{\rm eff}(\Lambda,\Omega, J_0)}{i\partial \Lambda_{\msvec n}} &=& 
 -\Omega_{\msvec n} + \left(D^{-1}[\Lambda] \right)_{\msvec n \msvec n} + H^2 \left(\sum_{\msvec m} \left(D^{-1}[\Lambda] \right)_{\msvec {mn}}\right)^2 = 0 , 
 \label{eq:lambda}\\
\frac{\partial S_{\rm eff}(\Lambda,\Omega, J_0)}{\partial \Omega_{\msvec n}} &=&  U^\prime(\Omega_{\msvec n} ) -i\Lambda_{\msvec n} = 0.
\label{eq:rho}
\end{eqnarray}
To obtain a solution to the second equation, we have to take $i \Lambda_{\msvec n}$ real at the saddle point.
Assuming a translation invariant solution, we thus take 
\begin{equation}
i \Lambda_{\msvec n}=i\Lambda_0\equiv 3+\frac{M^2}{2}= {\rm const.} \quad (M\ge 0), \qquad\quad \Omega_{\msvec n}=\Omega_0={\rm const.} ,
\label{spoint}
\end{equation}
with which eqs.  (\ref{eq:lambda}) and (\ref{eq:rho}) become
\begin{eqnarray}
\frac{H^2}{M^4} + I(M)&=& \Omega_0, \label{eq:saddle1} \\
 3+\frac{M^2}{2} &=& U^\prime(\Omega_0),
\label{eq:saddle2}
\end{eqnarray}
where
\begin{equation}
I(M) = \frac{1}{V}\sum_{\bf K} \frac{1}{\hat{\bf K}^2 + M^2}, 
\end{equation}
with
\begin{equation}
\hat{\bf K}^2 =\sum_\nu \hat{\rm K}_\nu^2,\quad \hat{\rm K}_\nu = 2\sin\frac{{\rm K}_\nu}{2} , \quad {\bf K}=2\pi\left(\frac{l_0}{L_0},\frac{l_1}{L_1},\frac{l_2}{L_2} \right)  , 
\end{equation}
and $l_\mu$'s are integers which satisfy $0\le l_\mu\le L_\mu-1$. 
Later we will solve these equations explicitly in detail.

As is well known \cite{MoZi,DaKeNe, DaKeNe2, KeNe}, it is possible to systematically organize 
the large $N$ expansion as perturbation theory around the saddle point, by introducing the quantum fluctuation 
as
\begin{eqnarray}
\Lambda_{\msvec n} &=& \Lambda_0 +\tilde \Lambda_{\msvec n} , \quad
\Omega_{\msvec n} = \Omega_0 + \tilde \Omega_{\msvec n},
\end{eqnarray}
 together with $J_{\msvec n}^i = J_0^i +\tilde  J_{\msvec n}^i$. 
In this paper we mainly consider up to 4-pt functions in the leading large $N$ expansion 
and for this purpose it is sufficient to consider Gaussian fluctuations of fields. (A brief discussion on the 6-pt function will be presented later.) 
Using the saddle point equations (\ref{eq:lambda}) and (\ref{eq:rho}), 
we expand the effective action as
\begin{eqnarray}
S_{\rm eff}(\Lambda,\Omega, J) &=& S_{\rm eff}(\Lambda_0,\Omega_0, J)
+ \sum_{\msvec n}\left[T_{\msvec n}(\tilde J) i\tilde \Lambda_{\msvec n} +\frac{1}{2}U^{\prime\prime}(\Omega_0) \tilde \Omega_{\msvec n}^2 -i\tilde\Lambda_{\msvec n}\tilde \Omega_{\msvec n}\right]\nonumber \\
&+&  \sum_{\msvec {nm}}\frac{1}{2} S_{\msvec {nm}}(J) \tilde \Lambda_{\msvec n} \tilde \Lambda_{\msvec m}+\dots,
\label{eq:expansion}
\end{eqnarray}
where
\begin{eqnarray}
T_{\msvec n}(\tilde J) &=& \frac{1}{N}\sum_i  \left[\left( G\,  \tilde J^i\right)_{\msvec n} \right]^2 
+\frac{2 H}{\sqrt{N} M^2} \left(G\,  \tilde J^N\right)_{\msvec n},  \\
S_{\msvec {nm}}(J) &=&  2 G_{\msvec{mn}}^2+\frac{4}{N} \sum_i \left( G\, J^i \right)_{\msvec n} G_{\msvec{nm}} \left( G\, J^i  \right)_{\msvec m} 
\end{eqnarray}
with
\begin{eqnarray}
G_{\msvec{nm}} &\equiv& (D^{-1}[\Lambda_0])_{\msvec{nm}}= \frac{1}{V}\sum_{\bf{K}}
G({\bf K}) {\rm e}^{i{\bf K} (\msvec{n}-\msvec{m}) } , \quad
G({\bf K}) = \frac{1} {\hat{\bf K}^2 +M^2}  .
\end{eqnarray}
We then perform the Gaussian integral for $\tilde \Omega_{\msvec n}$ and $\tilde\Lambda_{\msvec n}$, and finally obtain
\begin{eqnarray}
Z(J) &\simeq& {\cal N}_3 \exp[ - N W(J) ]
\label{eq:master}
\end{eqnarray}
where
\begin{eqnarray}
W(J) &=& S_{\rm eff}(\Lambda_0,\Omega_0,J) +\frac{1}{2}\sum_{\msvec{nm}} T_{\msvec n}(\tilde J) K^{-1}_{\msvec {nm}}(J)T_{\msvec m}(\tilde J) +\frac{1}{2N}{\rm Tr}\, \ln K(J) , 
\end{eqnarray}
with
\begin{eqnarray}
K_{\msvec{nm}}(J) &=& S_{\msvec{nm}}(J) +\frac{\delta_{\msvec{nm}}}{U^{\prime\prime}(\Omega_0)}, \\
S_{\rm eff}(\Lambda_0,\Omega_0,J) &=& V\left[ U(\Omega_0) - U^\prime(\Omega_0)\Omega_0 \right] -\frac{1}{2}{\rm Tr} \ln G -\frac{1}{2N} \sum_{\msvec{nm},i} J_{\msvec n}^i G_{\msvec{nm}} J_{\msvec m}^i ,
\\
{\cal N}_3 &=& {\cal N}_2\left(\frac{2\pi}{N\sqrt{U^{\prime\prime}(\Omega_0)}}\right)^{V} =\left(\frac{(2\pi)^N}{U^{\prime\prime}(\Omega_0)}\right)^{V/2} .
\end{eqnarray}

\subsection{VEV and Correlation functions}
We will call the first $N-1$ components of $\Phi$  "pions'' and
the $N^{\rm th}$ component "sigma", and denote $\Pi_{\msvec n}^i = \Phi^i_{\msvec n}$ for $i=1,2,\cdots, N-1$ and $\Sigma_{\msvec n} =\Phi^N_{\msvec n}$, respectively.
Taking derivatives of $W(J)$ in eq. (\ref{eq:master}) with respect to the source $J$ and setting $ J\rightarrow J_0$, 
we can calculate the leading large $N$ contributions to the connected part of arbitrary $n$-point functions.

We start with the vacuum expectation value (VEV) of $\Sigma_{\msvec n}$ as
\begin{eqnarray}
\sqrt{N}\Sigma &\equiv&  \langle \Sigma_{\msvec n}\rangle=\left.  \frac{-N\partial W(J)}{\partial J^N_{\msvec n}}\right\vert_{J\rightarrow J_0} = \sqrt{N} H \sum_{\msvec{m}}G_{\msvec{nm}} -\frac{1}{2}{\rm Tr}\, K^{-1}(J_0)\left.  \frac{\partial K(J)}{\partial J_{\msvec n}}\right\vert_{J\rightarrow J_0}, \nonumber \\
&=& \sqrt{N}\frac{H}{M^2}\left[1 -\frac{4}{N} \sum_{\msvec {lm}} K^{-1}_{\msvec{lm}}(J_0) G_{\msvec{ml}}G_{\msvec{mn}}\right]
= \sqrt{N}\frac{H}{M^2}\left[1 + O\left(\frac{1}{N}\right)\right] ,
\end{eqnarray}
where
\begin{eqnarray}
K_{\msvec{n m}}(J_0)&=&\frac{1}{V}\sum_{\bf K} {\rm e}^{i{\bf K}(\msvec{n}-\msvec{m})}\mathcal{K}({\bf K}),\\
\mathcal{K}({\bf K})&=&\frac{1}{U^{\prime\prime}(\Omega_0)}+2J({\bf K})+\frac{4H^2}{M^4}G({\bf K}), \\ 
J({\bf K}) &=& \frac{1}{V}\sum_{\bf Q} G({\bf Q}) G({\bf K}+{\bf Q}) .
\end{eqnarray}

The connected 2-pt function for $\Pi$ is given by
\begin{eqnarray}
\langle \Pi_{\msvec n}^i \Pi_{\msvec m}^j \rangle_c &\equiv& \left.  \frac{-N\partial^2 W(J)}{\partial J^i_{\msvec n}\partial J_{\msvec m}^j}\right\vert_{J\rightarrow J_0} = \delta^{ij}\left[
G_{\msvec{nm}} -\frac{4}{N} \sum_{\msvec{ls}} K^{-1}_{\msvec{sl}}(J_0) G_{\msvec{nl}}G_{\msvec{ls}} G_{\msvec{sm}}\right]\nonumber \\
&=&  \delta^{ij}\left[\frac{1}{V}\sum_{\bf K} G({\bf K}) {\rm e}^{i{\bf K}(\msvec{n}-\msvec{m})}
 + O\left(\frac{1}{N}\right)
\right] ,
\end{eqnarray}
so that the pole mass of pions in lattice units becomes
\begin{equation}
M_\pi = 2 \sinh^{-1} \frac{M}{2}.
\end{equation}
The connected 2-pt function for $\Sigma$,  on the other hand, can be obtained as
\begin{eqnarray}
\langle \Sigma_{\msvec n} \Sigma_{\msvec m} \rangle_c &\equiv& \left.  \frac{-N\partial^2 W(J)}{\partial J^N_{\msvec n}\partial J^N_{\msvec m}}\right\vert_{J\rightarrow J_0} = 
G_{\msvec{nm}} -\frac{4H^2}{M^4} \sum_{\msvec{ls}}G_{\msvec{nl}}K^{-1}_{\msvec{ls}}(J_0) G_{\msvec{sm}}
+ O\left(\frac{1}{N}\right),\nonumber \\
&=&\frac{1}{V} \sum_{\bf K} \Gamma^{(2)}_\Sigma({\bf K}) {\rm e}^{i{\bf K}(\msvec{n}-\msvec{m})}
 + O\left(\frac{1}{N}\right)
\end{eqnarray}
where
\begin{eqnarray}
 \Gamma^{(2)}_\Sigma({\bf K})   &=&\frac{G(\mvec{K})}{\mathcal{K}(\mvec{K})}
 \left[\mathcal{K}(\mvec{K})-\frac{4H^2}{M^4}G(\mvec{K})\right] .
\end{eqnarray}

The connected 4-pt function for $\Pi$ is evaluated as\footnote{Note that the lattice delta function we use here is periodic.}
\begin{eqnarray}
&&\langle \Pi_{\msvec{n}_1}^{i_1} \Pi_{\msvec{n}_2}^{i_2} \Pi_{\msvec{n}_3}^{i_3} \Pi_{\msvec{n}_4}^{i_4} \rangle_c \equiv \left.  \frac{-N\partial^4 W(J)}{\partial J^{i_1}_{\msvec{n}_1}\partial J^{i_2}_{\msvec{n}_2}\partial J^{i_3}_{\msvec{n}_3}\partial J^{i_4}_{\msvec{n}_4}}
\right\vert_{J\rightarrow J_0}\nonumber \\
&=& \frac{1}{V^3}\sum_{{\bf K}_1,{\bf K}_2,{\bf K}_3,{\bf K}_4}{\rm e}^{i({\bf K}_1\msvec{n}_1+{\bf K}_2\msvec{n}_2+{\bf K}_3\msvec{n}_3+{\bf K}_4\msvec{n}_4)}
\delta^{(3)}({\bf K}_1+{\bf K}_2+{\bf K}_3+{\bf K}_4) \frac{1}{N}\nonumber \\
&\times& G({\bf K}_1)G({\bf K}_2)G({\bf K}_3)G({\bf K}_4)    
 \left[ \delta^{i_1i_2}\delta^{i_3i_4}\left\{ \Gamma^{(4)}_{\Pi}({\bf K}_1+{\bf K}_2)   +O\left(\frac{1}{N}\right) \right\}+ 2{\rm \ perms\ }   \right],\nonumber \\
\end{eqnarray}
where
\begin{eqnarray}
 \Gamma^{(4)}_{\Pi}({\bf K}) &=& -4\mathcal{K}^{-1}({\bf K}).
\end{eqnarray}
Similarly the connected  3-pt function for $\Pi\Pi\Sigma$ is given by 
\begin{eqnarray}
&&\langle \Pi_{\msvec{n}_1}^{i_1} \Pi_{\msvec{n}_2}^{i_2} \Sigma_{\msvec{n}_3} \rangle_c \equiv \left.  \frac{-N\partial^3 W(J)}{\partial J^{i_1}_{\msvec{n}_1}\partial J^{i_2}_{\msvec{n}_2}\partial J^{N}_{\msvec{n}_3} } \right\vert_{J\rightarrow J_0}
=\frac{1}{V^2}\sum_{{\bf K}_1,{\bf K}_2,{\bf K}_3}{\rm e}^{i({\bf K}_1\msvec{n}_1+{\bf K}_2\msvec{n}_2+{\bf K}_3\msvec{n}_3)}
\nonumber \\
 &\times& \delta^{(3)}({\bf K}_1+{\bf K}_2+{\bf K}_3)
 \frac{1}{\sqrt{N}}G({\bf K}_1)G({\bf K}_2)G({\bf K}_3)
\left[ \delta^{i_1i_2}\Gamma_{\Pi\Pi\Sigma}^{(3)}({\bf K}_3)  +  O\left(\frac{1}{N}\right) \right] 
\end{eqnarray}
where
\begin{eqnarray}
 \Gamma^{(3)}_{\Pi\Pi\Sigma}({\bf K}) &=& \frac{H}{M^2}\Gamma^{(4)}_{\Pi}({\bf K}) =\Sigma\, \Gamma^{(4)}_{\Pi}({\bf K}) .
\end{eqnarray}

\section{Renormalization and continuum limits}
\label{cont}
Let us  consider the renormalization of this model in the leading large $N$ for the infinite volume case,
in order to take various continuum limits. 

In the ${\mvec L} \to\infty$ limit the lattice sums discussed in the 
previous section become integrals:
\begin{eqnarray}
I_\infty(M)&=&\int_{-\pi}^\pi\frac{\mathrm{d}^3{\bf K}}{(2\pi)^3}
\frac{1}{{\hat {\bf K}}^2+M^2}, \\
J_\infty({\bf K})&=&\int_{-\pi}^\pi\frac{\mathrm{d}^3{\bf Q}}{(2\pi)^3}
\frac{1}
{({\hat {\bf Q}}^2+M^2)\left[\left(\widehat{{\bf K}+{\bf Q}}\right)^2+M^2
\right]} .
\end{eqnarray}
For small mass $M \ge 0$ and small external momentum ${\bf K}$, we have
\begin{eqnarray}
I_\infty(M)&=&I_0-\frac{ M}{4\pi}+I_2 M^2 +O(M^3),\\
J_\infty({\bf K})&=&\frac{1}{4\pi\vert{\bf K}\vert}\,\arctan\left(\frac{\vert{\bf K}\vert}{2M }
\right)- I_2 +O (M, {\bf K}),
\end{eqnarray}
where constants $I_0$ and  $I_2$  are numerically calculated as $I_0=0.2527310\cdots$ and $I_2=-0.0121641\cdots$.

There are two phases at $H= 0$ in the infinite volume, the symmetric (SYM) phase where $M^2\not=0$ and $\Sigma=H/M^2=0$, and the broken (BRO) phase where $\Sigma\not=0$ and $M^2=H/\Sigma=0$.
Eqs.~(\ref{eq:saddle1}) and (\ref{eq:saddle2})  at $\Sigma = M = H=0$ determine the phase boundary as
$3=U^\prime(I_0) $.

We now solve the saddle point equations, eqs.~(\ref{eq:saddle1}) and (\ref{eq:saddle2}) at $H= 0$ near the continuum limit.
In the SYM case, eqs.~(\ref{eq:saddle1}) and (\ref{eq:saddle2}) become
\begin{eqnarray}
I_0 -\frac{M}{4\pi} +I_2 M^2+O(M^3) &=&\Omega_0, \quad 3+\frac{M^2}{2} = U^\prime (\Omega_0),
\end{eqnarray}
while in the BRO case,  they are given by
\begin{eqnarray}
I_0+\Sigma^2&=&\Omega_0, \quad 3 = U^\prime (\Omega_0).
\end{eqnarray}

As can be seen later, it is convenient to  define an effective 4-pt coupling constant $u_{\rm eff}$ as
\begin{equation}
U^{\prime\prime}(\Omega_0) = \frac{u_{\rm eff}}{12} a,
\label{eq:effective_coupling}
\end{equation}
 where $a$ is the lattice spacing. In the case of the $\phi^4$ model, $u_{\rm eff} = U/a = g_4$.
 We define the nonlinear $\sigma$ model limit as the $u_{\rm eff}\rightarrow\infty$ limit.
 
\subsection{Continuum limits in the symmetric phase}
We first consider the SYM case. Defining the renormalized pion mass $m_R$ as $M=m_Ra$ ($m_R\ge 0$)  with the lattice spacing $a$ and expanding $U^\prime$ around $I_0$ at $O(a^2)$, we obtain with eq.~(\ref{eq:effective_coupling})
\begin{eqnarray}
Z m_R^2  + \frac{u_{\rm eff}}{24\pi}  m_R   -r_2 &=& O(a), 
\end{eqnarray}
where we define
\begin{eqnarray}
Z&\equiv&1 - \frac{U^{\prime\prime\prime}(I_0)}{16\pi^2} \le 1, 
\end{eqnarray}
and make an additional tuning such that
$
2U^\prime(I_0) -6 = r_2 a^2 
$ .
We here restrict ourselves to the case that the effective 6-pt coupling $U^{\prime\prime\prime}(I_0)$ is non-negative.
As long as $r_2 \ge 0$,
we obtain 
\begin{equation}
m_R  = \frac{\sqrt{u_{\rm eff}^2 + 4 (24\pi)^2Z r_2}-u_{\rm eff}}{48\pi Z},
\end{equation}
and $r_2=0$ corresponds to the massless theory, while 
in the nonlinear $\sigma$ model limit we have
\begin{equation}
m_R  = \frac{24\pi  r_2}{u_{\rm eff}}
\label{eq:special}
\end{equation}
if  $r_2 =O(u_{\rm eff})$. 
Using $m_R$, the renormalized 2-pt function in the momentum space is given by
\begin{equation}
\gamma_\pi^{(2)}({\bf k}) =\frac{1}{{\bf k}^2+m_R^2} .
\end{equation}

Instead of the dimensional parameter $r_2$, we use the renormalize pion mass $m_R$,  in addition to the dimensional coupling $u_{\rm eff}$, in order to specify the continuum limit.  Then $r_2$ is expressed by others as
\begin{eqnarray}
r_2 &=& Z m_R^2\left( 1 + \frac{1}{24\pi Z \alpha }\right) \ge 0
\end{eqnarray}
in the symmetric phase, 
where we introduce the ratio $\alpha = m_R /u_{\rm eff}$.
We call a massive theory ($m_R\not =0$) case A and a massless theory ($m_R=0$) case B, while the nonlinear $\sigma$ model limit is named as case I and otherwise as case II, so that  there are four different continuum limits, IA, IB, IIA, IIB.

We next  define the renormalized 4-pt vertex function in the symmetric phase as
\begin{eqnarray}
\gamma^{(4)}_\pi ({\bf k}) &=& \frac{\Gamma_\Pi^{(4)}({\bf K} = {\bf k} a)}{ a}\rightarrow
-\frac{1}{3}\frac{u_{\rm eff}} {\displaystyle 1+\frac{u_{\rm eff}}{24\pi\vert{\bf k}\vert} \arctan\left(\frac{\vert{\bf k}\vert}{2 m_R}\right)},
\end{eqnarray}
where we use the effective coupling defined in  eq.~(\ref{eq:effective_coupling}).

\begin{figure}[tbh]
\begin{center}
\scalebox{0.4}{\includegraphics{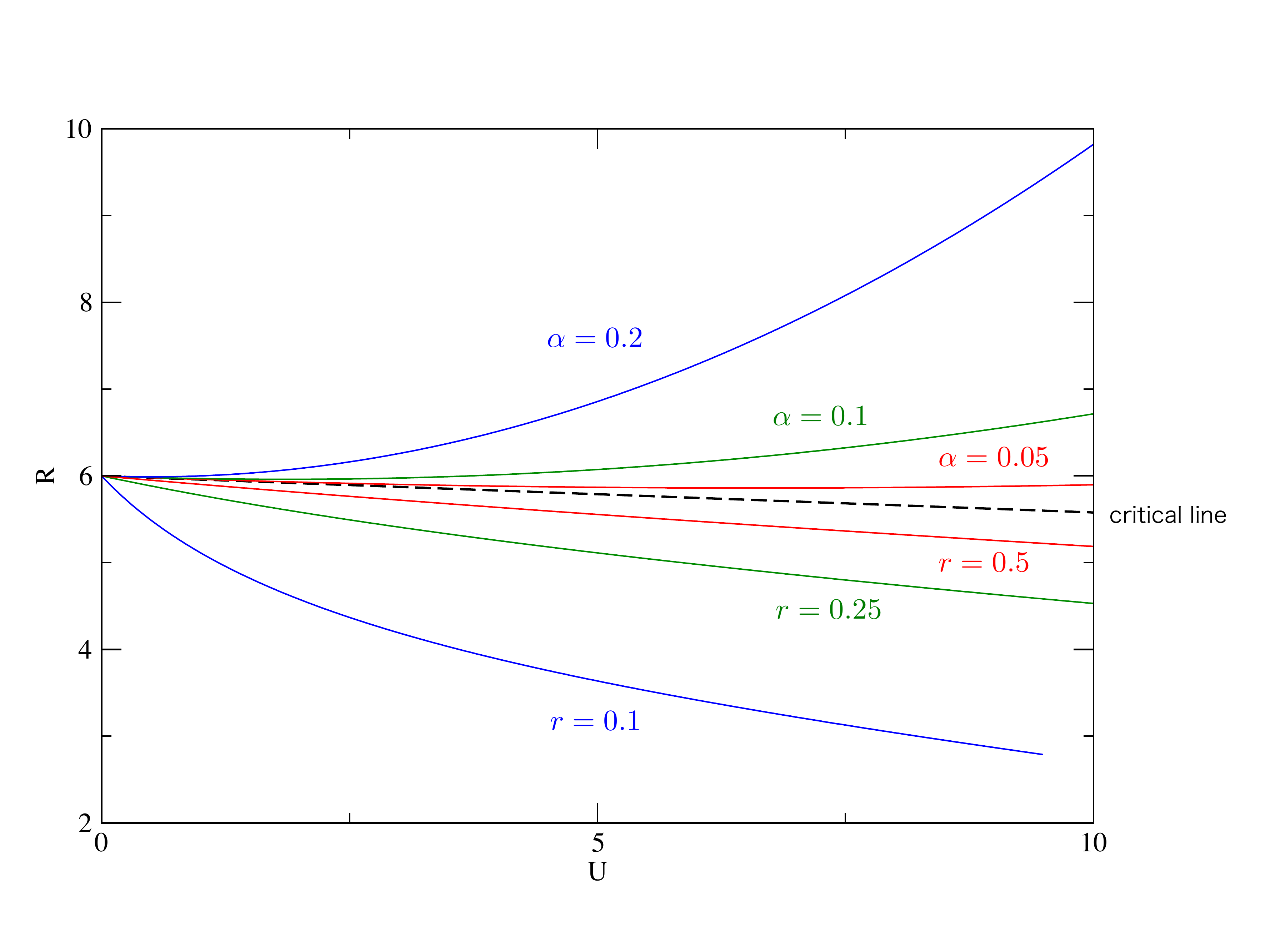}}
\end{center}
\caption{Lines of constant physics for $G=G_{k\ge 4}=0$. The dashed black line is the critical line. In the symmetric phase (above the critical line)
the red, green and blue curves are lines of constant $\alpha=0.05,0.1,0.2$ respectively. In the broken phase
the red, green and blue curves are lines of constant $r=0.5,0.25,0.1$
respectively.}
\label{fig:LCP}
\end{figure}

 We define a line of constant physics (LCP) by a curve  on which $ \alpha \equiv  m_R/u_{\rm eff} $ is kept constant even  at finite lattice spacing. Taking $G=G_{k\ge 4}=0$ for simplicity, the LCP is determined by
\begin{equation}
\frac{R}{2}=3 +\frac{\alpha^2 U^2}{2} -\frac{U}{12} I _\infty ( \alpha U) .
\end{equation}
In Fig.~\ref{fig:LCP}, we draw this LCP at several values of $\alpha$ in the $(U, R)$ plane.

\subsection{Continuum limits in the broken phase}
In the broken phase, using the saddle point equation in the $H\rightarrow 0$ limit, we obtain
the renormalized condensate $\sigma_R^2 =\Sigma^2/a$ in the continuum limit as
\begin{eqnarray}
\sigma_R^2\equiv w_R&=& \left\{
\begin{array}{ll}
\displaystyle \frac{ u_{\rm eff}}{12 Z_B}\left(1-\sqrt{1+ r_2 Z_B \left(\frac{12}{u_{\rm eff}}\right)^2}\right), & Z_B\equiv U^{\prime\prime\prime}(I_0) > 0 \\
\displaystyle -6\frac{ r_2}{u_{\rm eff}} & Z_B =0
\\
\end{array}
\right.  .
\end{eqnarray}
Therefore $r_2 $ must be negative in the broken phase.
We call the broken phase the case C, and there are IC and IIC depending on $u_{\rm eff}$.

The renormalized 4-pt vertex function in the continuum limit is given by
\begin{eqnarray}
\gamma^{(4)}_\pi ({\bf k} )&=& -\frac{1}{3} \frac{u_{\rm eff}}{\displaystyle 1+\frac{u_{\rm eff}}{48\vert{\bf k}\vert} +\frac{w_R u_{\rm eff}}{3 {\bf k}^2}}  ,
\end{eqnarray}
which is related to
the $\pi\pi\sigma$ 3-pt vertex function in the continuum limit as
\begin{equation}
\gamma^{(3)}_{\pi\pi\sigma} ({\bf k} ) \equiv  \frac{\Gamma^{(3)}_{\Pi\Pi\Sigma}({\bf K}={\bf k}a)}{a^{3/2}}
=  \sqrt{w_R} \gamma^{(4)}_\pi ({\bf k}).
\end{equation}

Finally the continuum limit of the connected 2-pt function for $\sigma$ is given by
\begin{eqnarray}
\gamma^{(2)}_\sigma({\bf k}) &=& a^2 \Gamma^{(2)}_{\Sigma}({\bf K}={\bf k}a) =\frac{\displaystyle 1+\frac{u_{\rm eff}}{48 \vert {\bf k}\vert}}{\displaystyle {\bf k}^2 +\frac{u_{\rm eff}}{48}\vert {\bf k}\vert +\frac{w_R u_{\rm eff}}{3}} ,
\end{eqnarray}
which has a pole at
\begin{equation}
\vert {\bf k}\vert = \gamma ( -1 \pm \sqrt{1-64\beta} ), \quad \gamma=\frac{u_{\rm eff}}{96}, \quad \beta = 48\frac{w_R}{u_{\rm eff}} ,
\end{equation}
while the pion is massless in this phase.
If we interpret $\vert {\bf k}\vert$ as $\sqrt{-{\rm k}_0^2+{\rm k}_1^2+{\rm k}_2^2}$ and taking ${\rm k}_0= m_\sigma - i\Gamma_\sigma$, ${\rm k}_1={\rm k}_2=0$, we obtain 
\begin{equation}
-{\rm k}_0^2 = -(m_\sigma^2-\Gamma_\sigma^2) + 2im_\sigma\Gamma_\sigma =
2 \gamma^2 \left[ 1 - 32\beta \mp\sqrt{1-64\beta}\right] .
\end{equation}
If $0 \le \beta < 1/64$,  $-k_0^2$ is real and positive, so that there is no pole in the propagator.
On the other hand, if $1/64 \le \beta$, we have the $\sigma$ resonance, whose mass and width are given by
\begin{equation}
m_\sigma^2 = \gamma^2 (64\beta -1), \quad \Gamma_\sigma^2 = \gamma^2 .
\end{equation}

In Fig.~\ref{fig:LCP} we have also plotted lines of constant $r\equiv\Gamma_\sigma/m_\sigma$ (the ratio of
resonance width over mass) for the lattice regularization with $G=G_{k\ge 4}=0$.
For technical details we refer to Appendix~\ref{app_resonance}.
At $U\sim10$ on the lines for $r =0.5, 0.25, 0.1,$
the values of the lattice mass $M_\sigma$ are $\sim 0.20,0.41,1.0$. Lattice artifacts for the ratio $\Sigma^2/M_\sigma$ are rather
small along these lines (in the region plotted),
only deviating maximally a few percent from the continuum limit
$\omega_R/m_\sigma=(r+1/r)/32$.

\subsection{6-pt vertex function for $\pi$ fields}
We can also calculate the 6-pt vertex function for $\pi$ fields, by considering the third order of quantum fluctuation 
in eq.~(\ref{eq:expansion}) as
\begin{equation}
U^{\prime\prime\prime}(\Omega_0)\sum_{\msvec n}\frac{\tilde\Omega_{\msvec n}^3}{3!}-\frac{4i}{3}
\sum_{\msvec{n}_1,\msvec{n}_2,\msvec{n}_3}G_{\msvec{n}_1{\msvec n}_2}G_{\msvec{n}_2{\msvec n}_3}
G_{\msvec{n}_3{\msvec n}_1}\tilde\Lambda_{{\msvec n}_1}\tilde\Lambda_{{\msvec n}_2}\tilde\Lambda_{{\msvec n}_3}
\end{equation}
at the leading order of the large $N$ expansion.

The connected 6-pt function for $\pi$ in the continuum limit is $O(1/N^2)$, and  is given in the momentum space as
\begin{eqnarray}
\delta^{(3)}(\sum_{i=1}^6{\bf k}_i )\prod_{i=1}^6 \gamma_\pi^{(2)}({\bf k}_i)&\times&\frac{1}{N^2}
\left[\left\{
\gamma_\pi^{(6),{\rm 1PI}}({\bf k}_{12},{\bf k}_{34})\delta_{i_1i_2}\delta_{i_3i_4}\delta_{i_5i_6} + \mbox{14 perms}\right\}  \right. \nonumber \\
&+& \left.\left\{ \gamma_\pi^{(6),{\rm 1PR} }({\bf k}_{12},{\bf k}_{123},{\bf k}_{56})\delta_{i_1i_2}\delta_{i_3i_4}\delta_{i_5i_6} + \mbox{89 perms} \right\} \right]
\end{eqnarray}
where ${\bf k}_{ij}={\bf k}_i+{\bf k}_j$, ${\bf k}_{ijk}={\bf k}_{ij}+{\bf k}_k$, and the 1-particle irreducible(1PI) 6-pt vertex function is given by
\begin{eqnarray}
\gamma_\pi^{(6),1PI}({\bf k},{\bf p} ) &=& \gamma_\pi^{(4)}({\bf k})  \gamma_\pi^{(4)}({\bf p}) \gamma_\pi^{(4)}({\bf k}+{\bf p})\left[\frac{U^{\prime\prime\prime}(I_0)}{(u_{\rm eff}/6)^3}+T({\bf k},-{\bf p})\right], \label{eq:1PI}\\
T({\bf k},-{\bf p}) &=& \int\frac{d^3 q}{(2\pi)^3}\gamma_\pi^{(2)}({\bf q}) \gamma_\pi^{(2)}({\bf q}+{\bf k})
 \gamma_\pi^{(2)}({\bf q}-{\bf p})\nonumber \\
 &=& \frac{1}{16\pi}\int_0^1dx \int_0^1dy  \frac{ y}{\left[m_R^2 + f(x,y,{\bf k,p})\right]^{3/2}}
\end{eqnarray}
 with $f(x,y,{\bf k,p})={\bf k}^2 xy(1-xy) +{\bf p}^2 y(1-y) +2 {\bf k}\cdot {\bf p}xy(1-y)$, while the 1-particle reducible (1PR) 6-pt vertex function becomes
 \begin{eqnarray}
\gamma_\pi^{(6),1PR}({\bf k},{\bf p},{\bf q}) &=&  \gamma_\pi^{(4)}({\bf k}) \gamma_\pi^{(2)}({\bf p})\gamma_\pi^{(4)}({\bf q}) .
\label{eq:1PR}
\end{eqnarray}
Note that the term in eq.~(\ref{eq:1PR}) and the second term in eq.~(\ref{eq:1PI}) are generated by the 4-pt vertex.
Since the theory is super-renormalizable without bare 6-pt coupling,  however, they  do not generate any UV divergences once $m_R$ (or $w_R$) and $u_{\rm eff}$ are made finite in the continuum limit. Therefore the effective 6-pt coupling $U^{\prime\prime\prime}(I_0)$ is not required to renormalize the theory, so that  it can take an arbitrary non-negative value including zero.

The contribution from $U^{\prime\prime\prime\prime}(I_0)$ to the 8-pt vertex function, on the other hand,  vanishes in the continuum limit, showing that it corresponds to the non-renormalizable coupling in the continuum theory.

\section{Running  four-point coupling}
\label{g4}
Our definition of the running coupling $g_4$ is
\begin{equation}
g_4({\cal E})\equiv -\frac{3\gamma^{(4)}_\pi (\vert {\bf k}\vert={\cal E})}{\cal E}.
\end{equation}
This means that we measure the dimensionful four-point vertex function in units
of the energy scale at which it is measured. Our results crucially depend
on this definition, which we however found natural. Also this is the definition
that gives identically vanishing beta function for the conformal model, as
required.

There are six different continuum limits, A (massive symmetric), B(massless symmetric) and C (broken) times
I (finite $u_{\rm eff}$) and II ( $u_{\rm eff}\to\infty$) .
The corresponding running  coupling is given as follows (with $m_R > 0$).
\begin{equation}
\begin{split}
{\rm case\  IA:\ \ \ \ \ }g_4({\mathcal E})&=
\frac{24\pi}{\arctan\left({\mathcal E}/2m_R\right)}, \qquad g_4(0)=\infty, \quad g_4(\infty) = 48\\
{\rm case\  IB:\ \ \ \ \ }g_4({\mathcal E})&=48\\
{\rm case\  IIA:\ \ \ \ \ }g_4({\mathcal E})&=
\frac{24\pi}{24\pi{\mathcal E}\alpha/m_R+
\arctan\left({\mathcal E}/2m_R\right)}, \qquad g_4(0)=\infty, \quad g_4(\infty) = 0\\
{\rm case\  IIB:\ \ \ \ \ }g_4({\mathcal E})&=
\frac{48}{1+48{\mathcal E}/u_{\rm eff}}, \qquad g_4(0)=48, \quad g_4(\infty) = 0\\
{\rm case\  IC:\ \ \ \ \ }g_4({\mathcal E})&=
\frac{48{\mathcal E}/w_R}{16+{\mathcal E}/w_R}, \qquad g_4(0)=0, \quad g_4(\infty) = 48\\
{\rm case\  IIC:\ \ \ \ \ }g_4({\mathcal E})&=
\frac{48{\mathcal E}/w_R}{16+{\mathcal E}/ w_R+48 {\mathcal E}^2/(w_R u_{\rm eff})},
\qquad g_4(0)=0, \quad g_4(\infty) = 0 
\end{split}
\end{equation}

We see that case IB is conformal: $g_4\equiv48$. Case IA is Ultra-Violet (UV) conformal,
the range of $g_4$ is from $\infty$ to $48$ as we move from Infra-Red (IR) to UV. For 
case IIA the range is from $\infty$ to $0$ and thus it is an UV Asymptotically Free (AF) model.
For IIB the range of $g_4$ is from $48$ to $0$, this limiting model is
UV AF and conformal in the IR. In the broken phase case, IC can be called 
UV conformal again since $g_4$ grows from $0$ to $48$ and finally for
the generic broken phase case IIC $g_4$ grows from $0$ to a $\beta$-dependent
maximum value at ${\mathcal E}^2=w_R u_{\rm eff}/3$ and then decreases again to $0$.

The generic case IIA is our toy model for
YM theory and IIB corresponds to the case where there is also an IR
fixed point. For IIA and small $\alpha$, the theory is \lq\lq walking''
in the region around $g_4\approx48$.

\subsection{Beta function}

We take the usual definition
\begin{equation}
\beta_4(g_4)={\mathcal E}\frac{\partial}{\partial{\mathcal E}}g_4({\mathcal E}).
\end{equation}
Let us first consider the symmetric case. The beta function is always negative 
here (except for the conformal case IB,
where it is identically vanishing). For IA we have
\begin{equation}
\beta_4(g_4)=-\frac{g_4^2}{48\pi}\sin(48\pi/g_4),
\label{betaIA}
\end{equation}
while for IIB
\begin{equation}
\beta_4(g_4)=-g_4+\frac{g_4^2}{48}.
\label{betaIIB}
\end{equation}

For the generic case IIA the beta function can be implicitly given by first 
solving
\begin{equation}
\frac{2}{\pi}\arctan\xi+b\xi=\frac{48}{g_4}
\end{equation}
for the variable $\xi$, where $b=96\alpha$. The beta function is then
\begin{equation}
\beta_4(g_4)=-g_4+\frac{g_4^2}{24\pi}\left(\arctan\xi-\frac{\xi}{1+\xi^2}
\right).
\label{betaIIA}
\end{equation}
For small coupling we have the expansion
\begin{equation}
\beta_4(g_4)=-g_4+\frac{g_4^2}{48}-\frac{\alpha g_4^3}{6\pi}+\dots
\end{equation}

\begin{figure}[tbh]
\begin{center}
\scalebox{0.38}{\includegraphics{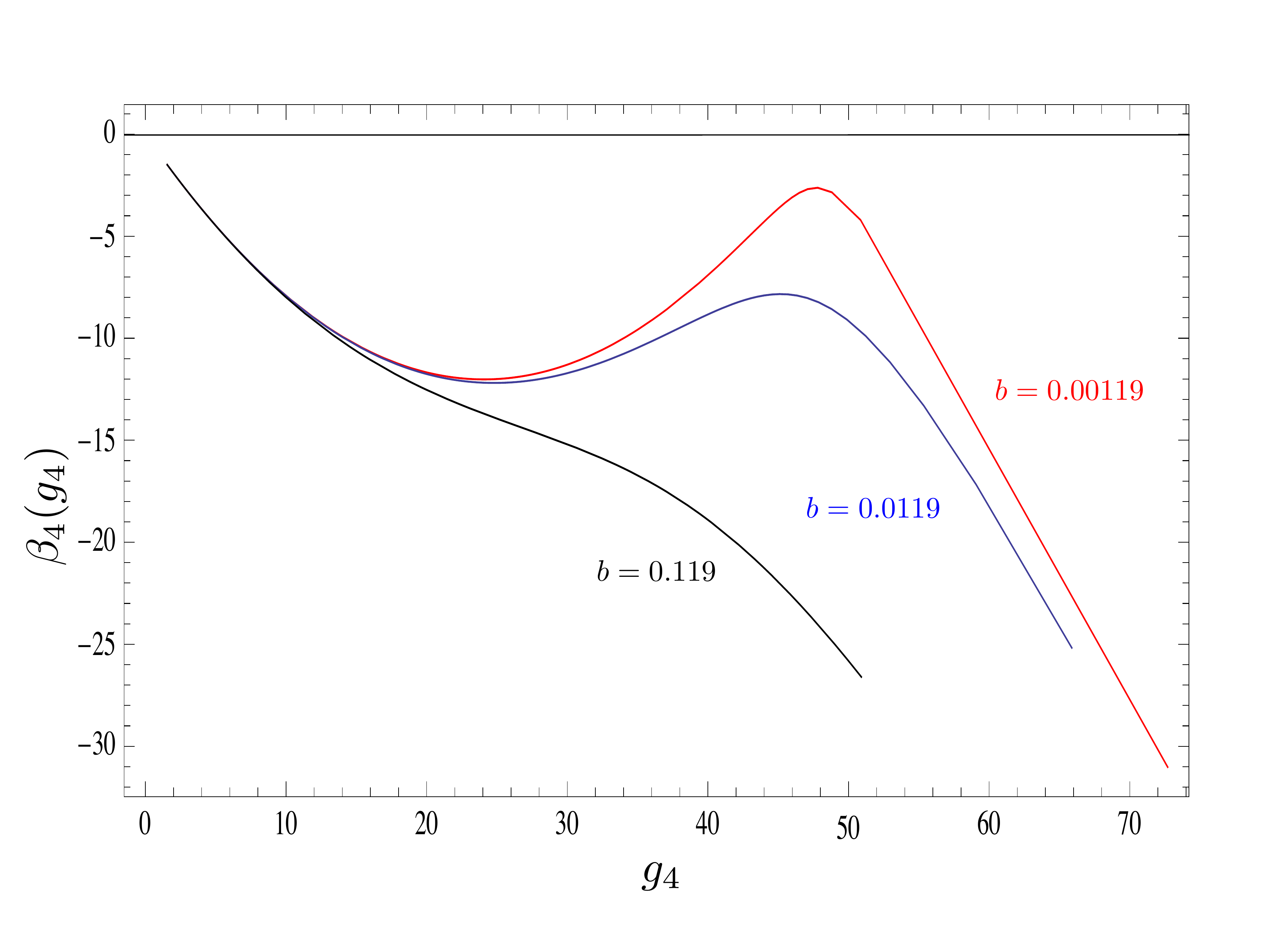}}
\end{center}
\caption{The beta function in the symmetric phase for $b=$ 0.119 (black), 0.0119 (blue) and 0.00119 (red). 
}
\label{fig:beta_cont}
\end{figure}

It is interesting to note that (\ref{betaIIB}) are the first two terms of
the weak coupling expansion of the beta function $\beta_4$ for any
$\alpha$ in case IIA.
For this generic case we can calculate the beta function numerically. It
depends on the parameter $b$. Fig.~\ref{fig:beta_cont} shows the beta function for
$b=96\alpha=$0.119, 0.0119  and 0.00119.
 They nicely show the walking behaviour for small $\alpha$: 
the beta function is close to (\ref{betaIIB}) for $0<g_4<48$,
and to (\ref{betaIA}) for $48<g_4<\infty$. 
We note that it is necessary to go extremely close to the conformal
point $\alpha=0$ to be able to observe \lq\lq walking" of the
beta function.

In the broken phase (in the generic case IIC) the beta function is 
a double valued function of the coupling:
\begin{equation}
\beta_4(g_4)=\pm\frac{g_4}{48}\sqrt{(48-g_4)^2-64\beta g_4^2}
=\pm\left\{g_4-\frac{g_4^2}{48}-\frac{\beta g_4^3}{72}+\dots\right\}.
\end{equation}
The coupling has finite range:
\begin{equation}
0\leq g_4\leq\frac{48}{1+8\sqrt{\beta}}
\end{equation}
and the beta function is
\begin{equation}
{\rm positive\ for\ \ \ }0<{\mathcal E}<\frac{4w_R}{\sqrt{\beta}},\qquad  
{\rm negative\ for\ \ \ }\frac{4w_R}{\sqrt{\beta}}<{\mathcal E}.
\end{equation}
For case IC the beta function is positive:
\begin{equation}
\beta_4(g_4)=g_4-\frac{g_4^2}{48}.
\label{betaIC}
\end{equation}

\subsection{Lattice artifact and finite size corrections to the "walking" behavior of the beta function}
In the case IIA, the running coupling in the continuum limit on the infinite volume is given by
\begin{eqnarray}
g_4({\cal E}) &=&  \frac{1}{\displaystyle x+\frac{1}{24\pi}\arctan\left(\frac{x}{2 \alpha}\right)}
\end{eqnarray}
where $x= {{\cal E}}/{u_{\rm eff}}$.
In this subsection, we consider an effect of non-zero $a$ and finite volume to the beta function.
For simplicity of analysis, we set $G=G_{k\ge 4}=0$, and then introduce the lattice spacing through $U$ as  $U = u_{\rm eff} a$. The running coupling thus becomes
\begin{eqnarray}
g_4^{\rm lat}({\bf K}) &=&  \frac{1}{\displaystyle x+\frac{x}{6} U J({\bf K} )}
\end{eqnarray}
where
\begin{eqnarray}
U J({\bf K} ) &=& \frac{U}{L_0L_1L_2} \sum_{l_{0,1,2}=0}^{L_{0,1,2}-1}\frac{1}{\left[ \widehat{\bf Q}^2 +\alpha^2 U^2\right] \left[\left(\widehat{{\bf K}+{\bf Q}}\right)^2 + \alpha^2 U^2\right]}\nonumber \\
&=& \prod_{i=0}^2\frac{1}{ L_iU} 
\sum_{l_{0,1,2}=0}^{L_{0,1,2}-1}
\frac{1}{\left[ \widehat{\bf q}^2 +\alpha^2\right] \left[ \left(\widehat{{\bf k}+{\bf q}}\right)^2 + \alpha^2\right]}, \\
{\bf Q} &=& 2\pi\left(\frac{l_0}{L_o},\frac{l_1}{L_1},\frac{l_2}{L_2}\right) ={\bf q} U, \quad
{\bf K} = 2\pi\left(\frac{n_0}{L_0},\frac{n_1}{L_1},\frac{n_2}{L_2}\right) ={\bf k} U
\end{eqnarray}
with ${\bf k}^2 = x^2$.

\begin{figure}[tbh]
\begin{center}
\scalebox{0.5}{\includegraphics{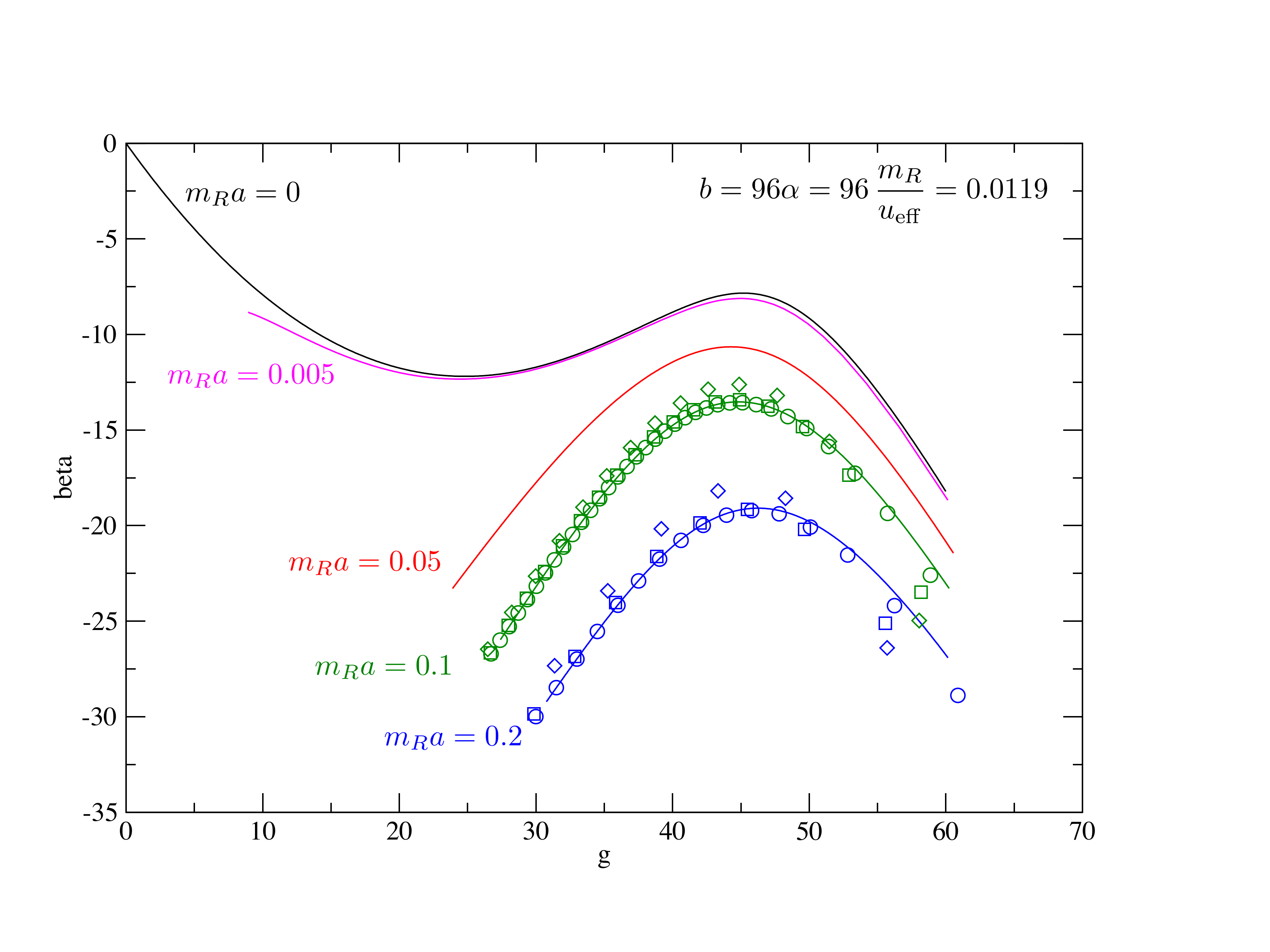}}
\end{center}
\caption{The beta function in the symmetric phase at $b=0.0119$ for several values of lattice spacings and the volume. For further details please consult the text.
}
\label{fig:beta_latt_FV}
\end{figure}

In Fig.~\ref{fig:beta_latt_FV},  we compare the beta function of $ g_4^{\rm lat}$ with that of $g_4$ as a function of $g_4$ in the symmetric phase at $b=0.0119$, where the "walking" behavior is clearly seen in the continuum limit (the black line). We take ${\bf K} = (K_0,0,0)$ with $ 0\le K_0 \le \pi$ in this calculation.
In the figure, the solid lines show the behavior of the beta function in the infinite volume  at finite lattice spacing, corresponding to $m_R a =$ 0.005 (magenta), 0.05 (red), 0.1 (green) and 0.2 (blue), where $U=96 \, m_R a/ b$.
As $a$ increases, the lattice beta function deviates from its continuum one, in particular at small $g_4$, where the energy scale $x$ becomes large due  to the asymptotic freedom.
We however still can observe the "walking behavior around $g_4\simeq 48$.
 In the finite volume case, we take $L_0=L_1=L_2=L$ for simplicity.
Instead of the derivative, we use the symmetric difference of the discrete energy $x$ in the finite volume to define the lattice beta function. 
In the  figure, symbols represent the beta function at $L =$ 30 (diamonds), 40 (squares) and 80 (circles)
at  $m_R a =$ 0.1 (green) while $L = $15 (diamonds), 20 (squares) and 40 (circles)
at  $m_R a =$ 0.2 (blue), which  give $m_R a L = $ 3,4 and 8 for both cases.
We observe that the finite size effect to the beta function is rather mild, except at strong coupling in the low energy region.

\section{Scattering phase shifts}
\label{scatt}
\subsection{Scattering amplitude}
The pion scattering amplitude in the large $N$ limit is given by
\begin{eqnarray}
T^{i_1i_2,i_3i_4}({\bf k}_1,{\bf k}_2\ \vert\  {\bf k}_3, {\bf k}_4) &\equiv & 
\lim_{{\bf k}_{1,2,3,4}\rightarrow  
\mbox{on-shell, } \sum_i {\bf k}_i = 0}
\nonumber \\
&&
\frac{1}{N-1}
\left[\delta^{i_1i_2}\delta^{i_3i_4} \gamma_\pi^{(4)}({\bf k}_1+{\bf k}_2) + \mbox{ 2 perms}\right] ,
\end{eqnarray}
where on-shell momenta in the center of mass system are given by ${\bf k}_1 =(iE_p,\vec p)$, ${\bf k}_2 =(iE_p,-\vec p)$, ${\bf k}_3 =(-iE_q,\vec q)$, ${\bf k}_4 =(-iE_q,-\vec q)$ with $E_p^2 =\vec{p^2} +m_R^2=E_q^2 =\vec{q^2}  +m_R^2$,  $\vec p=(p_1,p_2)$ and $\vec q=(q_1,q_2)$.
Explicitly we have
\begin{eqnarray}
T^{i_1i_2,i_3i_4}({\bf k}_1,{\bf k}_2\ \vert\  {\bf k}_3, {\bf k}_4) &= &
\frac{1}{N-1}\left[\delta^{i_1i_2}\delta^{i_3i_4} \gamma_\pi^{(4)}(2iE_p,\vec 0)\right. \nonumber \\
&+&\left.
\delta^{i_1i_3}\delta^{i_2i_4} \gamma_\pi^{(4)}(0,\vec p+\vec q) 
+\delta^{i_1i_4}\delta^{i_3i_2} \gamma_\pi^{(4)}(0,\vec p-\vec q) 
\right] .
\end{eqnarray}

In terms of the "isospin" decomposition for $N-1$ pions 
\begin{equation}
T^{i_1i_2,i_3i_4}({\bf k}_1,{\bf k}_2\ \vert\  {\bf k}_3, {\bf k}_4) =\sum_{I=0}^2 Q_I^{i_1i_2,i_3i_4}\, T_I(\vec p,\vec q)
\end{equation}
with projectors
\begin{eqnarray}
Q_0^{i_1i_2,i_3i_4} &=&  \frac{1}{N-1}\delta^{i_1i_2}\delta^{i_3i_4}\,, 
\\
Q_1^{i_1i_2,i_3i_4}&=&\frac12\left(\delta^{i_1i_3}\delta^{i_2i_4}-\delta^{i_1i_4}\delta^{i_2i_3}\right)\,,
\\
Q_2^{i_1i_2,i_3i_4}&=&\frac12\left(\delta^{i_1i_3}\delta^{i_2i_4}+\delta^{i_1i_4}\delta^{i_2i_3}\right)
-\frac{1}{N-1}\delta^{i_1i_2}\delta^{i_3i_4}\,,
\end{eqnarray}
we obtain
\begin{eqnarray}
T_0 (\vec p,\vec q) &=& \gamma_\pi^{(4)}(2iE_p,\vec 0)+\frac{1}{N-1} \left[\gamma_\pi^{(4)}(0,\vec p+\vec q) 
+ \gamma_\pi^{(4)}(0,\vec p-\vec q) \right]  , \\
T_1 (\vec p,\vec q) &=& \frac{1}{N-1} \left[ \gamma_\pi^{(4)}(0,\vec p+\vec q) 
-  \gamma_\pi^{(4)}(0,\vec p-\vec q) \right] , \\
T_2 (\vec p,\vec q) &=& \frac{1}{N-1} \left[ \gamma_\pi^{(4)}(0,\vec p+\vec q) 
+  \gamma_\pi^{(4)}(0,\vec p-\vec q) \right] .
\end{eqnarray}
Therefore, in the large $N$ limit, we have
\begin{eqnarray}
T_0 (\vec p,\vec q) &=& \lim_{\varepsilon\rightarrow 0}\gamma_\pi^{(4)}((i-\varepsilon)W,\vec 0), \qquad
T_1 (\vec p,\vec q) =
T_2 (\vec p,\vec q) = 0 ,
\end{eqnarray}
where $W= 2 E_p$.

Using the integral formula in the continuum limit
\begin{eqnarray}
\lim_{a\rightarrow 0} \lim_{\varepsilon\rightarrow 0} a J\left( (i-\varepsilon)W a,\vec 0\right) &=&\frac{1}{4\pi W}{\rm arccoth} \left(\frac{W}{2m_R}\right) + i\frac{1}{8W}, 
\end{eqnarray}
we have
\begin{eqnarray}
T_0 (\vec p,\vec q) &=& -\frac{1}{X+iY},
\label{eq:amp}
\end{eqnarray}
where
\begin{eqnarray}
X&=&\left\{
\begin{array}{ll}
\displaystyle  \frac{3}{u_{\rm eff}}+\frac{1}{8\pi W}{\rm arccoth} \left(\frac{W}{2m_R}\right), & \mbox { SYM } \\
& \\
\displaystyle \frac{3}{u_{\rm eff}}-\frac{w_R}{W^2}, & \mbox{ BRO}
 \end{array}
 \right. ,
 \qquad
Y= \frac{1}{16W} .
\end{eqnarray}

\subsection{Unitarity and scattering phase shift}
The scattering amplitude $T_0$ in eq.~(\ref{eq:amp}) satisfies unitarity 
\begin{eqnarray}
i\left[T_0 -T_0^\dagger\right] (\vec p, \vec q)&=& -\frac{1}{2W} \int \frac{{\rm d}^2 k}{(2\pi)^2}\frac{1}{2E_k}(2\pi)\delta(W-2E_k) T_0(\vec p, \vec k) T_0^\dagger(\vec k, \vec q), 
\end{eqnarray}
where $E_k^2=\vec{k^2} + m_R^2$.
Therefore, $T_0$ can be expressed as
\begin{equation}
T_0(\vec p,\vec q) = 16 W {\rm e}^{i\delta_0(W)}\sin \delta_0(W)
\end{equation}
where $\delta_0(W)$ is the scattering phase shift for the $I=0$ channel, so that
we obtain
\begin{eqnarray}
\cot \delta_0(W) &=& -\frac{X}{Y} =\left\{
\begin{array}{ll}
\displaystyle -\frac{48W}{u_{\rm eff}}-\frac{2}{\pi}{\rm arccoth} \left(\frac{W}{2m_R}\right), & \mbox{ SYM } \\
\\
\displaystyle -\frac{48W}{u_{\rm eff}}+\frac{16w_R}{W}, &  \mbox{ BRO } \\
\end{array}
\right. .
\end{eqnarray}

\begin{figure}[tbh]
\begin{center}
\scalebox{0.4}{\includegraphics{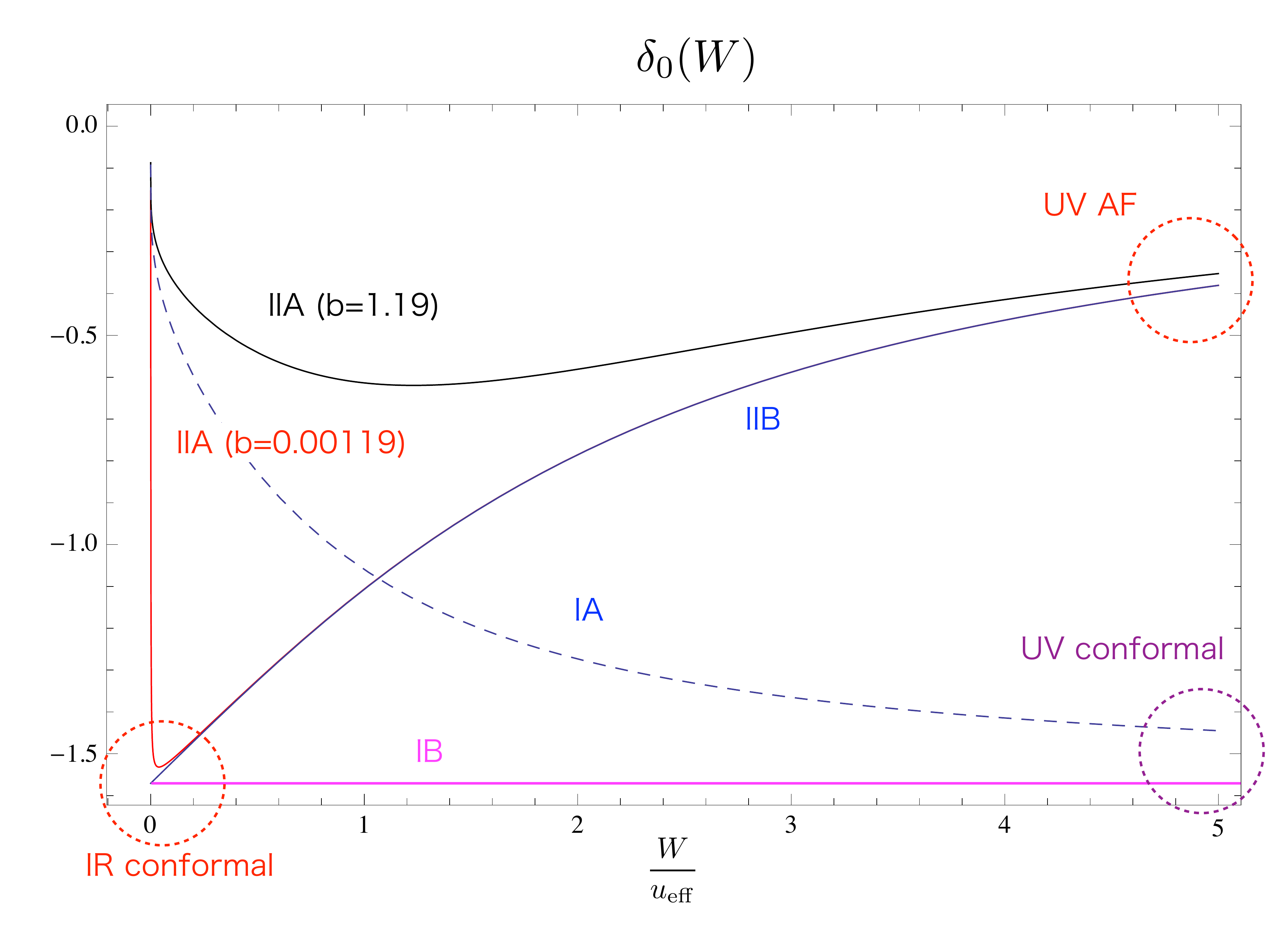}}
\end{center}
\caption{Scattering phase shift $\delta_0(W)$ as a function of $W/u_{\rm eff}$ for
IA(dashed line), IB(solid magenta line), IIA with $\b=96\alpha=1.19$ (solid black line), IIA with $b=0.00119$ (solid red line) and IIB (solid blue line).
}
\label{fig:ScatteringPhase}
\end{figure}

Fig.~\ref{fig:ScatteringPhase} shows behaviors of $\delta_0(W)$ as a function of $W$  in the symmetric phase, which clearly reflect behaviors of the running coupling in the various continuum limits:
In the case IA given by the dashed line, $\delta_0(W)$ starts from 0 at $W=0$ and monotonically approaches  $-\pi/2$ as $W$ increases. $\delta_0(W) = -\pi/2$ is the value for the conformal theory, as shown by the solid magenta line corresponding to the case IB. This shows that the case IA is UV conformal.
The behavior of $\delta_0(W)$ in the general case IIA depends on the mass, namely on the parameter $b=96\alpha = 96 m_R/u_{\rm eff}$.  At relatively large $m_R$ ($b=1.19$) denoted by the solid black line, $\delta_0(W)$ first decreases from 0 as $W$ increases but, at some value of $W$, it starts to increase  toward  0, showing that the IIA case is asymptotically free in the UV. If we decrease $m_R$ ( $b=0.00119$ ), $\delta_0(W)$ rapidly decreases from 0 to $-\pi/2$ (the conformal value) near $W=0$ and gradually increases toward 0 for increasing $W$, as shown by the solid red line. In the case of the massless limit, the case IIB, $\delta_0(W) = -\pi/2$ at $W=0$ and monotonically increases toward 0, showing the theory is IR conformal and UV asymptotically free. The IIA case with small mass such as $b=0.00119$ is nearly conformal, and therefore has "walking" coupling.

This shows that, although the running coupling is not a physical observable and depends on how it is defined, it captures some properties of the scattering phase shift. In other words, it opens a possibility to identify  a nearly conformal theory (or the walking coupling)  unambiguously from the physical observable, the scattering phase shift.  
   
\section{Summary}
\label{summary}
Due to the difficulties in establishing walking behavior
in candidate models in 4-dimensions using lattice simulations,
it is helpful to have simple toy models which show this 
phenomenon and where the problematic systematic errors 
can be investigated in detail. 
Although some features discovered in such models 
may be quite different in realistic models, 
such studies may give some valuable insights 
and indications where caution should be applied.

With this motivation we have shown that the O($N$) scalar field
theory in 3 dimensions exhibits walking behavior in the
large $N$ limit when a continuous parameter $\alpha=m_R/u_{\rm eff}$, 
the ratio of renormalized mass to an effective 
4-point coupling (with dimension mass), is very small. 
The characteristic walking, a slow change in dynamical
behavior over a range of (low) energies, has been
demonstrated for various couplings defined in infinite and 
finite volumes and also for the (on-shell) scattering phase shift,
as well as asymptotic freedom at high energies.

Some potential problems for lattice simulations
were identified. Firstly we have seen that 
(in the symmetric phase) for $\alpha$ in the range where 
walking sets in, it would, for infinite volume couplings,
be practically extremely difficult 
to reach energies where asymptotic freedom sets in. 
Recursive finite size scaling methods would certainly 
help in this respect.

Secondly although a qualitative characteristic of walking 
behavior, a maximum of the beta function at some non-zero 
value of the coupling, is observed at moderate correlation 
lengths, quantitative conclusions are marred by large 
lattice artifacts. The additional finite volume effects 
at medium energies are not so large and do not distort the results provided 
$m_RL\ge 3$. 

The toy model in the broken phase exhibits also rather 
interesting dynamics, which remains to be explored in detail.
In particular it manifests a resonance, and various 
continuum limits are characterized by values of $r=\Gamma_\sigma/m_\sigma$, 
the ratio of the width to the resonance mass.
Again here systematic sources of errors associated with the
lattice regularization can be studied.
For example along lines of constant physics defined
through keeping $r$ fixed, we have found that
lattice artifacts in the ratio of the vacuum expectation
value squared to the mass are rather small for correlation
lengths $>\sim2$.

\section*{Acknowledgement}
S. A. is supported in part by the Grant-in-Aid of the Japanese Ministry of Education, Sciences and Technology, Sports and Culture (MEXT) for Scientific Research (No. 25287046) and by MEXT Strategic Program for Innovative Research (SPIRE) Field 5 and Joint Institute for Computational Fundamental Science (JICFuS).
This investigation has also been supported by the European Union
and the State of Hungary and co-financed by the European Social Fund in the framework
of TAMOP-4.2.4.A/ 2-11/1-2012-0001 ${}^\prime$National Excellence Program$^\prime$.
S.A would like to thank the Wigner Research Center for Physics for its kind hospitality during his stay for this research project.
S. A and J. B. would like to thank the Max-Planck-Institut f\"ur Physik for its kind hospitality during their stay for this research project.

\appendix
\section{Resonance parameters on the lattice}
\label{app_resonance}
\newcommand{\uq}{\underline{Q}}

In this subsection we work in the broken symmetry phase 
with vanishing 6 or higher point couplings ($G=G_{k\ge4}=0$) and zero external field ($H=0$). 
The denominator of the lattice $\sigma$ propagator has the form:
\begin{equation}
\mathcal{D}(K)\equiv
\hat{K}^2\left(\frac{6}{U}+J_\infty(K)|_{M=0}\right)+2\Sigma^2\,.
\end{equation}
Setting $K=(K_0,\underline{0})\,\,,w=\hat{K}_0^2$ and performing the
$Q_0$ integration in eq.(3.2) we obtain:
\begin{equation}
\mathcal{D}(w)\equiv\mathcal{D}(K)|_{K=(K_0,\underline{0})}
=w\left(\frac{6}{U}+R(w)\right)+2\Sigma^2\,,
\end{equation}
where
\begin{equation}
R(w)=\int_{-\pi}^\pi\frac{\rmd^2\uq}{(2\pi)^2}
\frac{v(\uq)}{\hat{\uq}^2(\hat{\uq}^2+4)+w}\,,
\,\,\,\,\,\,
v(\uq)=\frac{\hat{\uq}^2+2}{\sqrt{\hat{\uq}^2(\hat{\uq}^2+4)}}
\end{equation}
$R(w)$ is analytic in $w$ with a cut starting at $w=0$ extending 
along the negative real axis to $w=-96$. It has the spectral
representation:
\begin{equation}
R(w)=\int_0^{96}\rmd s\,\frac{\rho(s)}{s+w}\,,
\end{equation}
with spectral function
\begin{equation}
\rho(s)=\frac{1}{4\pi^2\sqrt{s}}K\left(\sqrt{1-p(s)^2}\right)\,,
\end{equation}
where
\begin{equation}
p(s)=\frac14\left[6-\sqrt{4+s}\right]\,,
\end{equation}
and $K(k)$ is the elliptic function:
\begin{equation}
K(k)\equiv\int_0^{\pi/2}\rmd t\,\frac{1}{\sqrt{1-k^2\sin^2(t)}}\,.
\end{equation}

We are interested in identifying a sigma resonance 
for bare parameters close to the continuum limit 
i.e. we seek zeros of $\mathcal{D}(w)$ at small
$w=w^{(0)}$ on the second sheet:
\begin{equation}
w^{(0)}=-M_\sigma^2(1-ir)^2\,,
\end{equation}
where $r$ is the ratio of width to mass.
Noting that on the first sheet the discontinuity across the cut is
\begin{equation}
R(-s-i\epsilon)-R(-s+i\epsilon)=2\pi i\rho(s)\,.
\end{equation}
we can analytically continue to the second sheet (at least to a 
region near the cut) according to  
\begin{equation}
R^{(II)}(w)=R(w)+2\pi i\rho(-w)\,.
\end{equation}
The equations determining the resonance parameters are then
given by
\begin{eqnarray}
\frac{2\Sigma^2}{M_\sigma^2}&=&
\frac{(1+r^2)^2}{2r}\left[-\mathcal{D}(M_\sigma^2,r)
+2\pi\Re\rho(-w^{(0)})\right]\,,
\\
\frac{6}{U}&=&-\mathcal{C}(M_\sigma^2,r)+2\pi\Im\rho(-w^{(0)})
+\frac{(1-r^2)}{2r}\left[-\mathcal{D}(M_\sigma^2,r)
+2\pi\Re\rho(-w^{(0)})\right]\,,
\nonumber\\
&&
\end{eqnarray}
where
\begin{eqnarray}
\mathcal{C}(M_\sigma^2,r)&=&
\int_{-\pi}^\pi\frac{\rmd^2\uq}{(2\pi)^2}
 \frac{v(\uq)\left\{\hat{\uq}^2(\hat{\uq}^2+4)-M_\sigma^2(1-r^2)\right\}}
{\left\{\hat{\uq}^2(\hat{\uq}^2+4)-M_\sigma^2(1-r^2)\right\}^2
+4r^2M_\sigma^4}\,,
\\
\mathcal{D}(M_\sigma^2,r)&=&2rM_\sigma^2
\int_{-\pi}^\pi\frac{\rmd^2\uq}{(2\pi)^2}
\frac{v(\uq)}
{\left\{\hat{\uq}^2(\hat{\uq}^2+4)-M_\sigma^2(1-r^2)\right\}^2
+4r^2M_\sigma^4}\,.
\end{eqnarray}

\section{Finite-volume mass gap and running coupling}
\label{appB}
An interesting alternative running coupling is the LWW finite volume coupling\cite{Luscher:1991wu}
defined through the variable 
\begin{equation}
z=\ell m(\ell ),
\end{equation}
where $\ell = L a$ is the extension of the periodic box in physical units and
$m(\ell)$ is the particle mass when confined to the box, $m_R=m(\infty)$.
Our purpose here is to show the robustness of the walking behavior of our
model by demonstrating that the qualitative properties are the same whether
we consider the infinite volume four-point coupling and corresponding beta 
function (as we did in the main text) or similar objects associated to the 
finite volume mass gap. 
In addition, in MC measurements the lattice size is always finite and we can 
make a virtue of necessity and use a finite volume coupling to study walking 
behavior.   

Later we will also use the dimensionless volume variables
\begin{equation}
 L_u=\ell u_{\rm eff},\qquad\qquad L_m=\ell m_R, \qquad {\rm and} \qquad
L_w=\ell w_R.
\end{equation}

On the lattice, we consider the $T\to\infty$ limit keeping $L$ finite and $H\to0$.
We define
\begin{equation}
I(z,M(L))=\frac{1}{L^2}\int_{-\pi}^\pi\frac{\mathrm{d}K_0}{2\pi}
\sum_{l_1,l_2}
\frac{1}{{\hat {\bf K}}^2+M^2(L)}.
\end{equation}
Fixing
\begin{equation}
z=LM(L)
\end{equation}
this has small $M(L)$ expansion
\begin{equation}
I(z,M(L))=I_0+M(L) x_1(z)+ O (M^2(L)),
\label{expand}
\end{equation}
or written alternatively 
\begin{equation}
\frac{I(z,M(L))}{a}=\frac{I_0}{a}+m(\ell )\,x_1(z)+O (a),
\label{aexp}
\end{equation}
where we have used that $
M(L)=m(\ell)a
$.
The meaning of the terms in (\ref{aexp}) is
\begin{equation}
{\rm linear\ divergence\ }+{\rm \ physical\ part\ }
+{\rm \ cutoff\ effects}.
\end{equation}
The linear divergence ($I_0/a$) is the same as for infinite volume. 
The physical part of the result should be the same as for dimensional regularization where there is no linear divergence and gives the finite result:
\begin{equation}
x_1(z)=-\frac{1}{4\pi}+\frac{1}{4\pi z}J(z^2/4\pi).
\end{equation}
The definition of this function is 
\begin{equation}
J(v)\equiv \int_0^\infty\frac{\mathrm{d}t}{\sqrt{t^3}}\,{\rm e}^{-vt}\left[S^2
\left(1/t\right)-1\right],
\end{equation}
where
\begin{equation}
S(x)=\sum_{n=-\infty}^\infty{\rm e}^{-\pi n^2x}.
\end{equation}
$J$ is positive and monotonically decreasing to $0$. For small $z$ 
\begin{equation}
J(z^2/4\pi)\approx \frac{2\pi}{z}+{\rm const.}
\end{equation}
The function $x_1(z)$ decreases from $\infty$ to $-1/4\pi$ and has small $z$
behaviour
\begin{equation}
x_1(z)\approx \frac{1}{2z^2}+O (1/z).
\end{equation}
There is a unique zero of this function at some $z=z_*$. $x_1(z)$ is positive
for $0<z<z_*$ and negative for $z>z_*$. For later purposes we also note that
the function $zx_1(z)$ decreases from $\infty$ to $-\infty$ and $x_1(z)/z$
also decreases (from $\infty$ to $0$) for $0<z<z_*$.

\subsection{Continuum limit(s)}

From the first gap equation (\ref{eq:saddle1}) we get the expansion
\begin{equation}
\Omega_0=I_0+\omega_1 a+\omega_2 a^2+\dots
\label{Omegaexpand}
\end{equation}
with
\begin{equation}
\omega_1=\frac{zx_1(z)}{\ell}.
\end{equation}
Let us denote the analogous expansion coefficients for the infinite volume
theory by $\tilde\omega_i$. The leading coefficient is:
\begin{equation}
{\rm SYM:\ }\tilde\omega_1=-\frac{m_R}{4\pi},\qquad\quad
{\rm BRO:\ }\tilde\omega_1=w_R.
\end{equation}
In addition, eq.~(\ref{eq:effective_coupling}) leads to
\begin{equation}
U^{\prime\prime}(I_0) =\left(\frac{u_{\rm eff}}{12} - U^{\prime\prime\prime}(I_0)\tilde\omega_1\right)a\ .
\end{equation}

From the second gap equation (\ref{eq:saddle2}) we see that
\begin{equation}
\frac{R}{2}=3+\frac{M^2}{2} -\left( U^\prime(\Omega_0) - \frac{R}{2}\right) 
\label{3R}
\end{equation}
has to be volume-independent. Although we do not need the actual value
of couplings in the potential $U(S)$, their volume independence gives enough information
to calculate the mass gap in finite volume.

The volume independence leads to the equation,  
\begin{eqnarray}
\frac{U^{\prime\prime\prime}(I_0)}{2}(2\tilde\omega_1\omega_1-\omega_1^2) -\frac{u_{\rm eff}}{12}\omega_1 +\frac{z^2}{2\ell^2}&=&
\frac{U^{\prime\prime\prime}(I_0)}{2}\tilde\omega_1^2 -\frac{u_{\rm eff}}{12}\tilde\omega_1 +\frac{m_R^2}{2},
\end{eqnarray}
where the right-hand side is volume-independent, so is the left-hand side.
It is interesting to see that $U^{\prime\prime\prime}(I_0)$ dependence appears in the equation for finite volume qualities, while such a $U^{\prime\prime\prime}(I_0)$ dependence shows up only in the 6-pt vertex in the infinite volume, which is $1/N$ suppressed compared to the 4-pt vertex, as already discussed before.
For simplicity of the analysis, we set   $U^{\prime\prime\prime}(I_0)=0$ in the remaining of this appendix.

As before, the case I is obtained from the case II in the $u_{\rm eff}\to\infty$ limit.
We concentrate on the generic case II and we write (with $U^{\prime\prime\prime}(I_0)=0$) 
\begin{equation}
\frac{zx_1(z)}{6\ell}=\frac{w_R}{6}-\frac{m_R}{24\pi}+\frac{1}{u_{\rm eff}}
\left(\frac{z^2}{\ell^2}-m_R^2\right).
\end{equation}
This formula is valid for both phases if we note that $w_R=0$ in the symmetric
phase and $m_R=0$ in the broken phase. We now list the equation determining the
finite volume mass gap in all cases.

\begin{itemize}

\item case IA

$L_m=-4\pi zx_1(z)$

\item case IB

$x_1(z)=0$

\item IIA
 
$z^2/L_u^2-zx_1(z)/(6L_u)=\alpha^2+\alpha/24\pi$ \ \ \ \ or\ \ \ \ 
$\alpha z^2/L_m^2-zx_1(z)/(6L_m)=\alpha+1/24\pi$

\item IIB

$L_u=6z/x_1(z)$

\item IIC

$zx_1(z)=\displaystyle L_w+\frac{\beta z^2}{8 L_w}$

\item IC

$zx_1(z)=L_w$

\end{itemize}

From these formulas we find the following qualitative behaviours.

\begin{itemize}

\item case IA

$z$ goes from $z_*$ to $\infty$ as $L_m$ goes from $0$ to $\infty$.
For small $L_m$, $z\simeq z_*$ so the model is UV conformal, as found
before. For large $L_m$, $z\approx L_m$.

\item case IB

$z=z_*$ constant, the model is conformal.

\item case IIA

$z$ moves from $0$ to $\infty$ as $L_u$ (or $L_m$) changes from
$0$ to $\infty$, the model is UV AF. For small $L_u$, 
$z\approx(L_u/12)^{1/3}$ and for large $L_u$, $z\approx L_m=\alpha L_u$.

\item case IIB

$z$ moves from $0$ to $z_*$ as $L_u$ is changed from $0$ to $\infty$, it is
UV AF and IR conformal. For small $L_u$ it also behaves as 
$z\approx(L_u/12)^{1/3}$.

\item case IIC

For $0<L_w <z_1\sqrt{\beta/8}$, $z(L_w)$ is monotonically 
increasing from $0$ to $z_1$, where $z_1$ is defined by 
$x_1(z_1)=\sqrt{\beta/2}$. For small $L_w$, 
$z\approx(4L_w/\beta)^{1/3}$. This model is also UV AF.
For $L_w>z_1\sqrt{\beta/8}$, $z(L_w)$ is monotonically decreasing
and for large $L_w$, $z\approx1/(2 L_w)$.

\item case IC

$z(L_w)$ is monotonically decreasing, $z(0)=z_*$, the model is UV 
conformal. For large $L_w$, $z\approx1/(2L_w)$.

\end{itemize}

\subsection{Finite volume coupling}

\begin{figure}[tbh]
\begin{center}
\scalebox{0.4}{\includegraphics{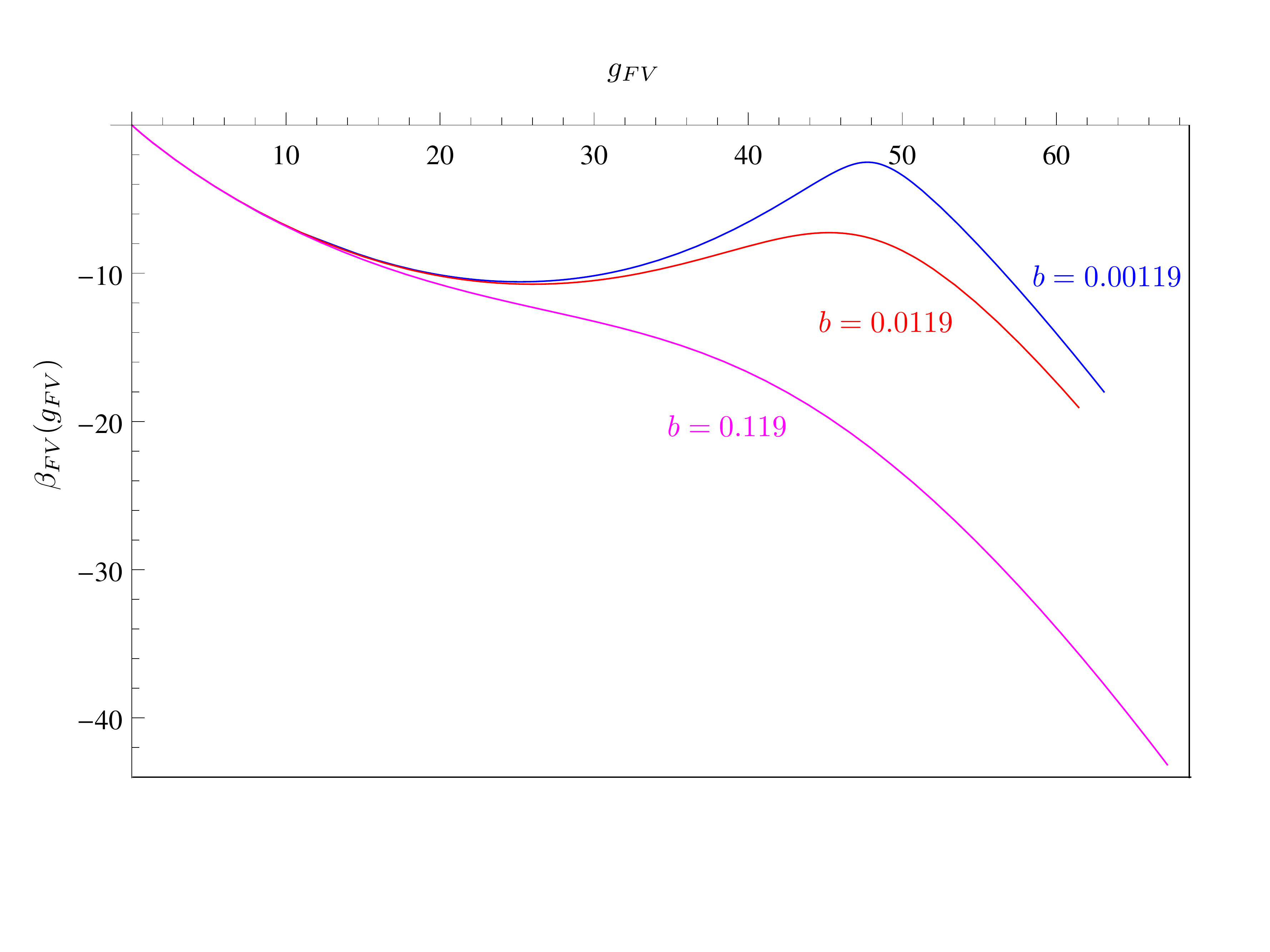}}
\end{center}
\caption{The beta function $\beta_{FV}(g_{FV})$ as a function of the finite volume coupling $g_{FV}$ at $b\equiv 96\alpha=0.119$(magenta line), 0.0119 (red line) and 0.00119 (blue line).
}
\label{fig:FV}
\end{figure}

Now we can define the finite volume running coupling $g_{FV}$ by
\begin{equation}
g_{FV}=\frac{48(z/z_*)^3}{p+(1-p)(z/z_*)^2},
\end{equation}
where $p$ is some constant. It is normalized to $48$ for the conformal points
and has qualitatively the same behaviour as $g_4$ and also the corresponding 
beta function
\begin{equation}
\beta_{FV}(g_{FV})=-L_u \frac{\partial}{\partial L_u}g_{FV}(L_u)
\end{equation}
shows walking behaviour for small $\alpha$.
Fig.~\ref{fig:FV} gives $\beta_{FV}(g_{FV})$ as a function of $g_{FV}$ at three values of $b$. 
Like the beta function of the 4-pt coupling in the text, the "walking" behavior in the finite volume coupling becomes more visible,  as $b$ decreases. 


  


\eject

\end{document}